\documentclass[conference]{IEEEtran}
\IEEEoverridecommandlockouts

\usepackage{cite}
\usepackage{amsmath,amssymb,amsfonts}
\usepackage{algorithmic}
\usepackage{graphicx}
\usepackage{textcomp}
\usepackage{hyperref}
\usepackage{subcaption}
\usepackage{xcolor}
\usepackage{tikz}
\usetikzlibrary{arrows.meta, positioning}
\usepackage{amsmath}

\def\BibTeX{{\rm B\kern-.05em{\sc i\kern-.025em b}\kern-.08em
    T\kern-.1667em\lower.7ex\hbox{E}\kern-.125emX}}
\begin{document}

\title{Optimizing Privacy-Preserving Primitives to Support LLM-Scale Applications
}

\author{\IEEEauthorblockN{Yaman Jandali, Ruisi Zhang, Nojan Sheybani, Farinaz Koushanfar}
\IEEEauthorblockA{University of California, San Diego \\
\{yeljanda,ruz032,nsheybani,farinaz\}@ucsd.edu
}
}

\maketitle

\begin{abstract}
Privacy-preserving technologies have introduced a paradigm shift that allows for realizable secure computing in real-world systems.
The significant barrier to the practical adoption of these primitives is the significant computational and communication overhead that is incurred when applied at scale. In this paper, we present an overview of our efforts to bridge the gap between this overhead and practicality for privacy-preserving learning systems using multi-party computation (MPC), zero-knowledge proofs (ZKPs), and fully homomorphic encryption (FHE). Through meticulous hardware/software/algorithm co-design, we show progress towards enabling LLM-scale applications in privacy-preserving settings. We show the efficacy of our solutions in several contexts, including DNN IP ownership, ethical LLM usage enforcement, and transformer inference.
\end{abstract}

\begin{IEEEkeywords}
Privacy-preserving machine learning, Hardware software co-design, Large language models
\end{IEEEkeywords}


\section{Introduction} 

Large Language Models (LLMs) continue to rapidly advance, demonstrating incredible capabilities across a wide variety of domains. In recent years, their utilization has reached beyond traditional natural language processing applications and into multiple critical areas such as software development, healthcare, and legal applications. Models such as Google's Gemini and OpenAI's GPT \cite{georgiev2024gemini, openai2025gpt5} have reached near expert performance in mathematical, coding, and question answering benchmarks \cite{chatterjee2025performance, georgiev2024gemini}. It is reasonable to consider a future where AI agents operate as assistants for humans, which may raise multiple concerns regarding their safety and security \cite{cui2024security, jain2023comprehensive}.

As LLM usage becomes more commonplace in sensitive domains, there is a fast growing need for privacy and verifiability for its safe deployment. Privacy-preserving technologies \cite{xu2021privacy} have emerged as technologies that enable efficient secure computing without revealing sensitive data. While existing work \cite{rechberger2022privacy} demonstrates the utilization of these fundamental cryptographic primitives to enable efficient privacy-preserving machine learning is possible, there is little research covering the unique challenges presented by LLMs. Such challenges include substantially higher computational demands, more frequent user interactions, and the handling of multi-modal inputs and outputs. This paper addresses this gap by proposing a co-design approach that jointly optimizes algorithms and systems to mitigate the computational and communication bottlenecks associated with scaling privacy-preserving technologies to LLM workloads.

In this paper, we begin by introducing the cryptographic techniques relevant to private LLM usage, including three core primitives, multi-party computation (MPC) \cite{yao1986}, zero-knowledge proofs (ZKPs) \cite{goldwasser1989knowledge}, and fully homomorphic encryption (FHE) \cite{gentry}. We outline their fundamental principles, strengths, and suitability for different parts of the LLM computation pipeline, emphasizing how each can perform secure computation without exposing sensitive data. Next, we focus on making MPC practical for LLM-scale inference by reducing its computational and communication overhead. We review recent advances such as hybrid MPC protocols (e.g., Chameleon~\cite{chameleon}), low-bitwidth and binarized model quantization (e.g., XONN~\cite{xonn}, COINN~\cite{coinn}), automated MPC circuit generation (e.g., MPCircuits~\cite{mpcircuits}), and crypto–ML co-design strategies that jointly optimize network architecture and cryptographic execution to improve scalability while maintaining model accuracy.

\begin{figure}[!htbp]
\centering
\includegraphics[width=0.95\columnwidth]{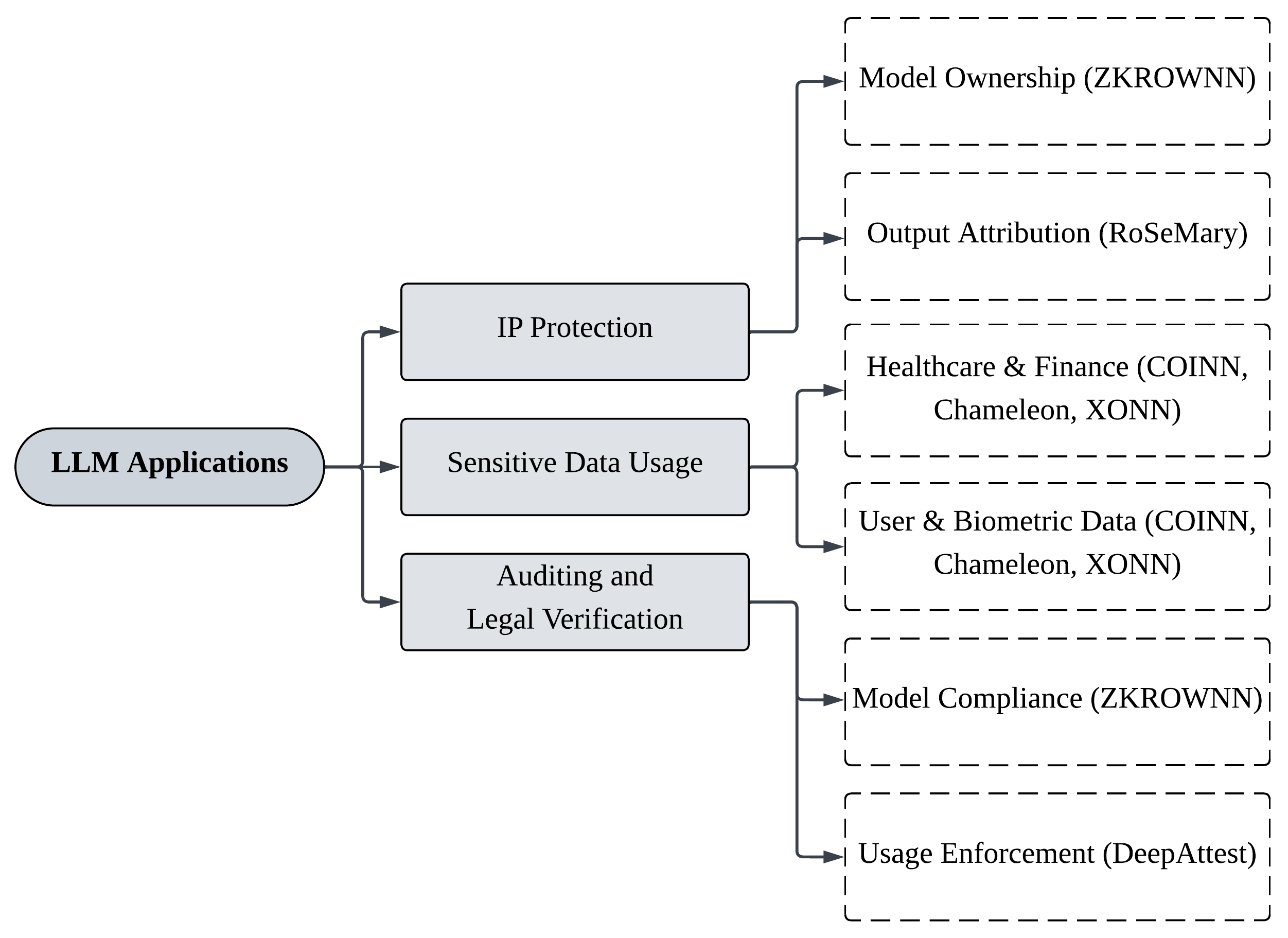}
\caption{Taxonomy of privacy-preserving LLM applications discussed in this paper. We outline three promising domains: IP Protection~\cite{zkrownn,rosemary}, Sensitive Data Usage~\cite{coinn,chameleon,xonn}, and Auditing and Legal Verification~\cite{zkrownn,deepattest}.}
\label{fig:taxonomy}
\end{figure}

Besides, we outline a few methodologies to ensure verifiability and traceability in privacy-preserving LLMs. It includes LLM-generated content watermarking~\cite{rosemary} and model ownership proof~\cite{zkrownn} through zero-knowledge proofs without revealing secret triggers. Finally, we explore future directions for scaling secure LLM deployment through crypto–hardware–algorithm co-design. As shown in Figure~\ref{fig:taxonomy}, we also outline promising application domains, such as auditing, legal verification, and sensitive data processing, where these 
fields would benefit from the enablement of robust and large-scale privacy-preserving AI systems.

\section{Overview of Cryptographic Techniques}


Large Language Model (LLM) inference involves a series of computations steps including (i) quantization~\cite{quant1, quant2}: reducing the numerical precision of model parameters/activations to reduce computation overhead, (ii) matrix multiplications~\cite{vaswani2023attentionneed, matMul2}: the majority of operations within attention and feed-forward layers, (iii) non-linear activation functions such as ReLU~\cite{reluMobileBert}, GELU~\cite{llama}, and Softmax~\cite{flashAttention} that are applied element-wise. In a privacy-preserving setting, each of these operations must be performed without revealing any sensitive data. To that end, we discuss three cryptographic techniques in order to support privacy preserving LLM pipeline.


\subsection{Multi-Party Computation}
Secure Multi-Party Computation (MPC) allows multiple parties to compute a joint function over their inputs while keeping each others’ inputs private. More formally, MPC enables $n$ parties, $P_1, \dots, P_n$, each holding a private input $x_i$, to jointly compute a function $f(x_1, \dots, x_n)$ to obtain an output $y = f(x_1, \dots, x_n)$, such that no party learns any information about the other parties' inputs beyond what is revealed by the output itself. This can be represented as:
$$ (y_1, \dots, y_n) \leftarrow f(x_1, \dots, x_n) $$
where each party $P_i$ receives their respective output share $y_i$ (or the full output $y$ if it's public).

MPC protocols are generally based on secret sharing and Yao's garbled circuits. Secret sharing schemes - such as in the protocols of Goldreich, Micali, and Wigderson - split data into shares which are distributed among parties such that the data can only be reconstructed when a sufficient number of shares - defined by a reconstruction threshold - are combined \cite{gmw}. For instance, in additive secret sharing, a secret $S$ can be split into $n$ shares $s_1, \dots, s_n$ such that $S = \sum_{i=1}^n s_i$. More generally, in Shamir's Secret Sharing \cite{shamir}, the secret is the constant term of a polynomial of degree $t$, and each party receives a share $(i, P(i))$, allowing reconstruction with any $t+1$ shares. This method of private computation enables secure yet efficient computation of linear operations. Yao's garbled circuits (GC) allows for the evaluation of boolean functions in a privacy preserving setting, which is particular suited for non-linear and branching operations \cite{yao1986}.

In practice, modern MPC protocols combine secret sharing (for arithmetic operations) and garbled circuits (for boolean/non-linear operations) to leverage the advantages of each. This method of hybrid design allows us to take advantage of MPC's strengths: linear operations over secret shares incur minimal overhead, and only the relatively smaller number of non-linear gates require the heavier cryptographic cost of GC.
When applying MPC to LLM inference, with transformer layers that involve large matrix multiplications, the scale of these operations magnifies the costs.

\subsection{Zero-Knowledge Proofs}

Zero-Knowledge Proofs (ZKPs) are a cryptographic protocol that allows a prover $\mathcal{P}$ to convince a verifier $\mathcal{V}$ that a given statement is true without revealing any information beyond just the validity of the statement itself. Within the domain of machine learning, ZKPs enable verifiable computation over private data, model parameters, or embedded signatures, making them well-suited for applications such as ownership proofs and usage governance for sensitive LLM applications.

A ZKP relies on three key properties: completeness, soundness, and zero-knowledge \cite{goldwasser1989knowledge}. \textit{Completeness} ensures that, for a given true statement, an honest prover can convince an honest verifier of the statement's validity. \textit{Soundness} guarantees that if the statement is false, a dishonest prover cannot successfully convince the verifier. \textit{Zero-knowledge} entails ensuring that the verifier learns nothing more than the truth of the statement itself. 

ZKPs can be interactive, requiring multiple rounds of communication between the prover and verifier, or non-interactive, in which case the prover sends a single message. This primitive has been extended to the field of verifiable machine learning, Zero-Knowledge Machine Learning (ZKML), where it can be used for privacy-preserving training, secure model verification, and authenticating AI-generated content \cite{ganescu2024trust, peng2025survey}. A specific class of non-interactive ZKPs, known as Zero-Knowledge Succinct Non-Interactive Arguments of Knowledge (zk-SNARKs), is particularly relevant for these applications due to its small proof size and rapid verification time \cite{ben2019aurora, groth2016snark}.

\subsection{Fully Homomorphic Encryption}
Fully Homomorphic Encryption (FHE) allows for computation over encrypted data directly, producing a still-encrypted output which, upon decryption, yields the intended result of the desired computation.

The idea of computing on encrypted data without decryption was first proposed by~\cite{HE}. The breakthrough of \textit{fully} homomorphic encryption (FHE) was introduced with Gentry's work in~\cite{gentry}, which presented the first scheme capable of supporting both additions and multiplications. Prior schemes had been only "partially" homomorphic, supporting either additions or multiplications but not both. 

In an FHE setting, a party encrypts its data and sends it to an untrusted second party. That party performs a computation on the encrypted input and returns the encrypted result, which only the original party can decrypt. Throughout the process, the computing party learns nothing about the original input or intermediate values.

Formally, an FHE scheme consists of four algorithms:
\begin{itemize}
    \item $\text{KeyGen}(\lambda) \rightarrow (pk, sk)$: Generates a public key $pk$ and a secret key $sk$ based on a security parameter $\lambda$.
    \item $\text{Encrypt}(pk, m) \rightarrow c$: Encrypts a plaintext message $m$ using the public key $pk$ to produce a ciphertext $c$.
    \item $\text{Decrypt}(sk, c) \rightarrow m$: Decrypts a ciphertext $c$ using the secret key $sk$ to recover the plaintext message $m$.
    \item $\text{Eval}(pk, f, c_1, \dots, c_k) \rightarrow c_f$: Takes the public key $pk$, a function $f$, and ciphertexts $c_1, \dots, c_k$ (encrypting $m_1, \dots, m_k$ respectively), and outputs a ciphertext $c_f$ such that $\text{Decrypt}(sk, c_f) = f(m_1, \dots, m_k)$.
\end{itemize}

The core property is that for any efficiently computable function $f$, if $c_1 = \text{Encrypt}(pk, m_1)$ and $c_2 = \text{Encrypt}(pk, m_2)$, then:
\[
\text{Eval}(pk, \text{add}, c_1, c_2) = \text{Encrypt}(pk, m_1 + m_2)
\]
\[
\text{Eval}(pk, \text{mult}, c_1, c_2) = \text{Encrypt}(pk, m_1 \times m_2)
\]
This enables arbitrary computations over encrypted data without ever decrypting it.



\begin{table*}[!ht]
    \centering
    \caption{A comparison of privacy-preserving machine learning techniques}
    \label{tab:comparison}
    \resizebox{\textwidth}{!}{%
    \begin{tabular}{|l|p{3.5cm}|p{3.5cm}|p{3.5cm}|p{3.5cm}|}
        \hline
        \textbf{Technique/Paper} & \textbf{Quantization} & \textbf{Matrix Multiplication} & \textbf{Activation Functions} & \textbf{Communication \& \newline Pre-processing} \\
        \hline
        XONN \cite{xonn} & Binarized to 1-bit & BNN-style fast matrix multiplication & Bit-sliced GC ReLU & Constant-round secure inference \\
        \hline
        COINN \cite{coinn} & Low-bitwidth quantization & Secret-shared matrix multiplication with reuse & GC-friendly QNN activations & Weight sharing, reduced GC cost \\
        \hline
        HELiKs \cite{heliks} & --- & Optimized FHE matrix multiplication & --- & Ciphertext packing, bootstrapping pipeline \\
        \hline
        Chameleon \cite{chameleon} & Fixed-point truncation and \newline clipping & Efficient Beaver triple matrix multiplication & ReLU / MaxPool via GC & MPC + GC hybrid efficiency \\
        \hline
        HE-Transformer \cite{he_transformer} & Reduced bit-width for internal data & Homomorphic matrix multiplication & Non-polynomial functions via approximation & Uses an FHE framework (e.g., SEAL) \\
        \hline
        CrypTFlow \cite{cryptflow} & Fixed-point arithmetic & Beaver triples based matrix multiplication & ReLU, MaxPool, etc. via MPC & MPC compiler for TensorFlow \\
        \hline
    \end{tabular}}
\end{table*}


\section{Making MPC Practical for LLMs}

In order to achieve efficient LLM-scale inference, we must overcome the computation and communication bottlenecks we face with multi-party computation (MPC). Key characteristics and comparisons of these techniques are summarized in Table \ref{tab:comparison}. We address this by optimizing matrix multiplications, and in this section, we highlight a series of innovations on these fronts which demonstrate that MPC is capable of scaling to the workload demands of LLMs. 

\subsection{Optimizing Matrix Multiplication with Hybrid MPC}




\textit{Chameleon}~\cite{chameleon} introduced a mixed-protocol framework that combines garbled circuits (GC) with additive secret sharing, leveraging the strengths of each approach. In Chameleon, the linear layers of a neural network (e.g., matrix multiplications and convolutions) are executed using lightweight arithmetic over secret shares, while the non-linear operations (such as activations like ReLU or comparison gates) are evaluated using Yao’s two-party garbled circuits. This hybrid design avoids the heavy symmetric-key cryptographic costs that a pure GC approach would incur for linear operations - which dominate deep learning workloads - and confines the expensive cryptographic work to the relatively smaller number of non-linear gates.

To enable two-party secure inference without requiring an online third party, Chameleon employs an offline pre-processing phase that generates large quantities of correlated randomness. This pre-processing can be done by a trusted dealer or via an initial setup protocol, and it allows almost all heavy cryptographic computations to be performed \textit{before} the actual inference. As a result, the online phase involves minimal communication and computation, drastically reducing the latency when the model is queried.

Chameleon supports fixed-point arithmetic to represent real-valued network parameters and activations with high fidelity, which is crucial for deep learning tasks. In benchmarks, a 5-layer convolutional neural network evaluated via Chameleon ran up to 133x faster than the CryptoNets scheme~\cite{cryptonets} and about 4.2x faster than the prior hybrid approach MiniONN~\cite{minionn} on the same task. While the original Chameleon work focused on modest-sized CNNs, its architecture-agnostic approach for mixing protocols is not limited to any specific network type. In fact, the ability to offload large linear layers to efficient secret-sharing makes this approach well-suited for transformer models, where matrix multiplications (in self-attention and feed-forward layers) dominate the computation. Notably, feed-forward networks (FFNs) account for a large portion of the parameters and FLOPs in transformer blocks~\cite{vaswani2023attentionneed, tay2022efficienttransformerssurvey}, making them an ideal target for MPC-friendly execution. We expect that Chameleon’s layered hybrid strategy can be directly extended to LLMs, laying important groundwork for secure inference on transformer-based models.

\subsection{Reduce Computation Overhead with Quantization}

\textit{XONN}~\cite{xonn} takes a complementary approach by adapting the model itself to better suit cryptographic evaluation. It targets \textit{binarized neural networks (BNNs)}, where both weights and activations are constrained to $\{-1, +1\}$ and thus can be encoded with single-bit precision. This extreme quantization allows expensive arithmetic operations in deep layers to be replaced by simple binary logic. In particular, XONN transforms each matrix multiplication into a series of XNOR operations, which incur essentially no cryptographic cost in Yao’s garbled circuits thanks to the Free-XOR optimization. By constraining the network architecture to rely mainly on XNOR and bit-counting operations, XONN manages to evaluate the linear layers with near-zero overhead in the GC domain.

To make this feasible for real-world models, XONN introduces a network customization pipeline that can retrain or fine-tune standard neural networks into their binarized counterparts while maintaining acceptable accuracy. It also provides a high-level API and compiler that translate models defined in common frameworks (e.g., Keras) into optimized garbled circuits, greatly streamlining secure deployment. Another key benefit of XONN’s all-GC approach is that it achieves constant-round secure inference: the number of communication rounds does not grow with the number of layers. This eliminates the traditional “interaction-per-layer” pattern that often hinders MPC scalability in deep networks.

Empirically, XONN achieves significant speedups over prior MPC frameworks. It runs oblivious inference up to 7x faster than Gazelle~\cite{gazelle}, 37x faster than SecureML~\cite{secureml}, and 93x faster than MiniONN~\cite{minionn} on comparable neural network benchmarks. Moreover, XONN was the first to demonstrate oblivious inference on a deep network as large as the 21-layer \textit{FitNet} CNN~\cite{fitnet}, marking a new scalability milestone for private deep learning. 



While XONN was primarily evaluated on vision CNNs, its core ideas carry implications for NLP and transformer models as well. Many recent efforts on efficient LLM deployment explore quantization down to 1--4 bits for weights and activations~\cite{quant1, quant2}. XONN demonstrates that pushing this to 1-bit not only reduces conventional inference costs but can also virtually eliminate cryptographic overhead for linear layers in secure inference. In this sense, XONN serves as an important proof-of-concept for applying aggressive low-bitwidth quantization to privacy-preserving LLM inference in the future.

\subsection{Automated and Scalable MPC Circuit Generation}

As deep learning models grow in size and complexity, \textit{manually} constructing optimized Boolean circuits for secure evaluation becomes impractical. \textit{MPCircuits}~\cite{mpcircuits} addresses this challenge by providing an end-to-end automated toolchain that generates MPC-optimized circuits from high-level descriptions. It adapts logic synthesis techniques from hardware design to translate functions - including entire neural networks - into Boolean circuits that are tailored for efficient secure computation.

A key innovation in MPCircuits is its awareness of the asymmetric costs of gates in cryptographic protocols. For example, in garbled circuits and similar MPC protocols, XOR (or XNOR) gates are “free” to evaluate under the Free-XOR technique, whereas AND gates (or other non-linear gates) incur most of the computational and communication cost. MPCircuits’ optimizer uses this knowledge to aggressively reduce the number of non-XOR gates in the generated circuit. This optimization of the logical representation of the computation results in as many operations as possible being carried out with XORs, leaving a minimal number of expensive operations. The result is a circuit that computes the correct function but much more efficiently under MPC, without requiring the developer to handcraft any low-level optimizations.

Another strength of the framework is its generality: MPCircuits supports both 2-party and $n$-party secure computation (such as the BMR protocol) by outputting circuits compatible with different back-end protocols. The system has been evaluated on a variety of benchmark applications - including secure auctions, voting, and neural network inference - often showing up to 4.2x runtime improvement compared to using unoptimized circuits or previous circuit-generation methods. By automating the circuit design process, MPCircuits significantly lowers the barrier to deploying large, complex models under MPC. A developer can start with a high-level description of a neural network and obtain an optimized Boolean circuit for it, without manual tuning or expertise in circuit design.

This capability is particularly important when considering components of modern LLMs, such as attention mechanisms and feed-forward layers, which can involve millions of gates when represented as circuits. Optimizing these at scale is a daunting manual task. By bridging the gap between hardware-oriented circuit synthesis and secure computation, MPCircuits makes it feasible to compile entire transformer models into MPC-efficient circuits, thereby enabling scalable privacy-preserving inference on increasingly large neural networks.

\subsection{Crypto-ML Co-Design}

\textit{COINN} is an end-to-end framework that pushes MPC-based DNN inference to new scales by co-optimizing the model architecture and the cryptographic execution \cite{coinn}. Rather than treating the network as fixed, COINN jointly adjusts how the neural network is quantized and structured alongside the choice of MPC protocols, minimizing the overall Boolean circuit complexity while maintaining accuracy. It introduces a domain-specific low-bitwidth quantization scheme that dramatically minimizes the bit-length of values throughout the model, directly reducing the cost of both linear and non-linear MPC operations. COINN also identifies opportunities to factor and reuse computations in secure matrix multiplications: by deliberately introducing weight sharing within each layer, many repeated multiplications can be replaced by cheaper additions which are low-cost in Boolean circuits. A custom two-party protocol based on oblivious transfer then multiplies these factored weights with the inputs without leaking the underlying shared values or their positions. This crypto/ML co-design enables logic synthesis optimizations in order to produce a substantially smaller and more efficient MPC circuit.

To make these optimizations practical at scale, COINN provides an automated compiler-like toolchain that assists in Boolean circuit generation for deep learning. Given a high-level model description (e.g., a PyTorch network), COINN’s design configurator automatically searches for the best per-layer quantization levels and factorization settings that balance accuracy and MPC cost. It leverages a genetic algorithm to efficiently traverse this large design space, using a combined score of model accuracy and secure execution cost to guide the selection of each layer’s parameters. The result is a tailor-made Boolean circuit (or equivalent MPC representation) for the entire DNN that contains minimal expensive operations (AND gates or arithmetic multiplies) and maximizes the use of cheap XOR additions. COINN’s toolchain then seamlessly deploys this optimized circuit with a mix of arithmetic sharing for linear layers and garbled circuits for non-linear layers, ensuring that each part of the computation is executed with the most efficient protocol. By integrating model adjustments with compiler-assisted circuit generation in this way, COINN produces secure inference pipelines that are far more performant than prior one-size-fits-all solutions.

Empirical results highlight COINN’s contributions to scalable MPC. In a two-party LAN setting it achieves 4.7x - 14.4x faster inference than the prior state-of-the-art on equivalent networks. Notably, these speedups come without an accuracy penalty - in fact, COINN slightly improves accuracy (by up to 4.7\%) compared to earlier MPC-tailored models, even as it slashes runtime by an order of magnitude. This balanced co-design enabled COINN to be the first to demonstrate oblivious inference on very large deep networks, including CNNs over 100 layers deep and involving billions of operations for ImageNet-scale tasks. Such a feat was previously out of reach for purely manual or homogeneous approaches. The focus on reusing logic and optimizing bitwidths is particularly pertinent to LLM-scale inference, where naively converting a transformer with hundreds of layers into a Boolean circuit would be infeasible. By tuning the network itself to be MPC-friendly, COINN reduces the circuit size and depth enough to make privacy-preserving inference on complex models tractable. In this way, COINN serves as a complementary piece in the broader effort to scale up secure MPC for real-world machine learning workloads: it bridges the gap between high-level model design and low-level circuit efficiency, ensuring that even as models grow to LLM proportions, their secure evaluation can remain within practical bounds.





\section{Verifiability and Ownership in Private LLMs}\label{watermarking_label}

Given the valuable nature of large scale LLMs~\cite{liu2024deepseek,bai2023qwen,dubey2024llama,team2024gemini} as intellectual property, there is strong drive for mechanisms to be put into place which would be able to reliably attribute and verify ownership of both models themselves and their generated outputs. These models are increasingly applied to critical domains such as software development, financial services, and scientific discovery, where the generated content often represents significant intellectual property. Ensuring that LLM outputs are source-verifiable and that their distribution can be traced in a tamper-resistant manner is therefore essential for protecting ownership rights and enabling accountable usage.

However, designing high-quality watermarks that meet robustness and usability requirements remains challenging—particularly for low-entropy content such as code and medical data. In many cases, watermark verification requires model owners to reveal their secret signatures, and reusing watermarked content often necessitates re-encoding, which can degrade usability. Cryptographic primitives such as zero-knowledge proofs offer a promising solution by enabling ownership verification without disclosing watermark details. In this section, we highlight two complementary approaches that integrate cryptographic techniques~\cite{zkrownn, deepattest} and neural network watermarking~\cite{darvish2019deepsigns,chen2019deepmarks,zhang2024emmark} to ensure verifiability without compromising privacy and utility.


\subsection{Zero-Knowledge Proofs of LLM-generated Content}
Beyond model ownership, watermarking can also enforce usage control over LLM-generated content. A prime example is AI-generated source code, where unregulated use could violate software licenses or introduce security risks. Simply detecting that code was machine-generated is not enough: we need a mechanism to attribute code back to the AI model and owner, so that unauthorized use can be flagged or prevented. This calls for robust watermarking of code outputs and a way to verify those watermarks without undermining the code’s usability. However, watermarking code is particularly challenging as code has a low entropy structure (limited places to hide a signal without altering functionality) and strict syntax/semantic requirements. Early methods such as inference-time token restriction for text (green-list decoding) prove inadequate for code, as they can break syntax or logic. Likewise, simple static transformations (e.g. renaming variables according to a secret mapping) are easily removed by adversaries.

\textit{RoSeMary} addresses these challenges with an ML/cryptography co-design that embeds ownership signatures into code in a resilient yet function-preserving manner \cite{rosemary}. It leverages a pre-trained code model (CodeT5 \cite{codet5}) to insert watermarks at the semantic level: the network learns to make subtle syntax or identifier modifications that encode a hidden bitstring while ensuring the program’s behavior remains unchanged. A corresponding neural decoder extracts the signature from a given code snippet. The encoder–decoder are trained jointly on many code examples (with random code perturbations during training) to optimize the classic triad of detectability, fidelity, and robustness. This yields high-quality watermarked code that preserves the original functionality while having a watermark with high detectability.

To enable verifiable usage control, RoSeMary integrates zero-knowledge proof verification. When a piece of code is suspected to originate from an LLM, a neutral arbiter can ask the model owner to prove the code contains their secret watermark – without the owner revealing the watermark or the code’s secrets. RoSeMary accomplishes this by having the owner provide the hidden signature as a private input to a ZKP circuit that runs the watermark extraction on the published code snippet. The result is a publicly verifiable proof that the code carries the owner’s signature, achieved without disclosing the signature or needing to alter the code after distribution. This secure verification remains efficient: RoSeMary’s prototype reports sub-second proof generation and verification (e.g. ~120 ms to verify a snippet) using modern zkSNARK libraries \cite{zksnark}. By binding each distributed code fragment with an invisible yet attestable signature, and using cryptographic proofs to check those signatures, RoSeMary empowers AI code generators to enforce proper use of their outputs. Model owners can thus confidently release assistive code to users, knowing that any downstream misuse or IP violations can be detected and proven without ambiguity.

\subsection{Zero-Knowledge Proofs of Model Ownership}

While watermarking neural networks \cite{chen2019deepmarks, darvish2019deepsigns, zhang2024emmark} provides a means for owners to mark their models, proving such ownership to a third party presents an additional challenge. As discussed in \textit{section \ref{watermarking_label}}, DNN watermarks are often composed of a signature within an output. In a legal or competitive setting, the model owner must demonstrate the watermark’s presence without revealing the secret trigger that may have been used to generate the watermark or watermark itself, since disclosure would enable adversaries to remove or forge it. Zero-knowledge proofs (ZKPs) offer a cryptographic solution: the owner can prove knowledge of a valid watermark trigger that causes the model to output their signature, without exposing the trigger or any model details. 

\textit{ZKROWNN} implements the first end-to-end framework for DNN ownership proof using zero-knowledge succinct non-interactive arguments of knowledge (zkSNARKs) \cite{zkrownn}. The core idea is to arithmetize the model’s watermark verification process, the forward pass on a secret input and checking of the signature, into a cryptographic circuit. During a one-time setup, the model prepares a zkSNARK proving key for the circuit representing the watermark verification logic, which is then used by the prover to generate a proof attesting to the watermark's presence, which any verifier can check with a public key. At dispute time, the owner (prover) uses the secret trigger as a witness to generate a succinct proof attesting to the watermark’s presence, which any verifier can check with a public key. Crucially, the proof reveals nothing about the trigger or the model beyond the claim of ownership’s validity. 

ZKROWNN demonstrates that such proofs can be made highly efficient in practice: a single proof ($\sim$ 128 bytes in size) is generated once and then publicly verifiable by anyone in under a second. Communication overhead is minimal (only a few kilobytes plus the constant-sized proof), and verification is non-interactive. This approach enables a third-party arbiter or client to confirm model ownership with strong cryptographic assurances. By combining watermarking with ZKPs, model owners gain a privacy-preserving way to enforce their IP rights, they can confidently prove a suspect model is a pirated copy of theirs without divulging the secret that marks it. 


Due to the low communication and verification overhead, the primitives ZKROWNN built for DNN models can potentially be extended to LLM scale with the same underlying computation components. By co-designing the watermarking algorithm with these primitives, researchers can explore efficient frameworks for these LLM applications.

\section{Future Directions}






\subsection{Crypto-Hardware-Algorithm Co-Design}

\subsubsection{Algorithm Redesign for Cryptographic Goals}
Meeting cryptographic security goals in AI systems can often require altering or augmenting the model itself to enforce protection. One promising approach is to redesign algorithms such that model usage is cryptographically restricted to authorized devices. For example, a model can be \textit{bound} to specific hardware through techniques like hardware fingerprinting and Trusted Execution Environments (TEEs). This method is demonstrated by \textit{DeepAttest}, which was the first to enable on-device DNN attestation by embedding a device-specific fingerprint in the model's weights and using a TEE to verify this fingerprint \cite{deepattest}. Under this framework, only a genuine model with the correct hidden fingerprint is able to execute on authorized hardware, preventing unauthorized copies or tampering even if the model or keys are leaked. Future AI deployments can build on this idea. For instance, with LLM usage increasingly pushed to cloud platforms, one could distribute encrypted models that only decrypt and run within secure enclaves on certified servers-thus enforcing controlled usage at the hardware level. Such cryptography-based attestation techniques provide a path toward robust model ownership protection and authorized-use enforcement in real-world deployments.

\subsubsection{Leveraging Existing Hardware for Secure Computation}
Beyond controlling model usage, leveraging hardware support wisely can drastically increase the efficiency of privacy-preserving computation. Offloading cryptographic workloads to specialized secure hardware helps overcome the performance bottlenecks of purely software solutions. This is showcased by \textit{Chameleon}, which uses a one-time trusted hardware setup to pre-generate correlated randomness for MPC, reducing the online inference phase to just a few lightweight steps \cite{chameleon}. This work demonstrates that even a limited amount of hardware trust can break through major bottlenecks faced in two-party secure inference. Similarly, \textit{Stamp} takes advantage of existing hardware by incorporating a small dedicated security processor to offload and accelerate the non-linear operations in MPC, substantially reducing the protocol’s overhead \cite{stamp}. These examples show that building secure components onto today’s hardware can yield significant speedups for privacy-preserving AI, making otherwise prohibitive secure computations more practical.

\subsubsection{Co-Designing Hardware with Cryptography in Mind}
In addition to using existing enclaves and processors, researchers are now exploring custom hardware accelerators built specifically for cryptographic workloads \cite{nvidia2023confidential, butt2024ifzkp, gu2025unizk, zhu2024szkp}. The goal of this line of work is to bake privacy directly into the next generation of AI chip design. For example, \textit{MPCircuits} introduces a hardware-aware circuit synthesis toolchain for MPC that optimizes boolean circuits for secure computation. By leveraging custom logic libraries tuned for cryptographic gates, this approach significantly decreases the number of expensive operations (e.g., AND gates) required for a given secure function, thereby reducing runtime. Likewise, \textit{HELiKs} (HE Linear Algebra Kernels) improves the throughput of homomorphic encryption by providing fast, hardware-optimized routines for encrypted matrix multiplication and convolution, drastically speeding up these operations while also decreasing communication overhead. Such co-design efforts illustrate how specialized hardware and algorithmic optimizations can jointly improve the performance of secure computation. By shrinking costs at the circuit level and accelerating encrypted primitives, cryptography-oriented hardware designs are bringing LLM-scale privacy-preserving computation closer to reality.

\subsubsection{Towards an End-to-End Heterogeneous Secure Hardware Pipeline}
Looking further ahead, we can envision a future in which cryptography and hardware are co-designed at every level of the stack, resulting in an end-to-end pipeline of heterogeneous secure hardware components. Considering the above advancements, one can imagine specialized neural network chips that natively support operations on encrypted data (e.g., homomorphic ciphertexts or MPC shares) at the scale of LLM computations. These would operate alongside other trusted components in a seamless architecture. For instance, an end-to-end system might employ TEEs for secure pre-processing and key management, dedicated MPC accelerators for multi-party inference, and homomorphic encryption co-processors for encrypted matrix operations, all orchestrated together. In such a design, data remains protected throughout the computation pipeline, and each hardware unit is optimized for its security task. Overall, crypto-hardware-algorithm co-design will be vital for future LLM-scale applications: by synergistically combining cryptographic techniques with hardware-level enforcement and acceleration, we can build AI systems that are both highly secure and high-performing. 

\subsection{Applications}

\subsubsection{Auditing and Legal Verification}

Large language models are increasingly used in high-stakes scenarios, motivating exploration into thorough auditing and legal verification mechanisms. Researchers have called for systematic LLM audits spanning multiple levels - from the model provider to downstream applications - to ensure systems remain ethical, legal, and technically robust. In practice, this means future LLM deployments may include transparent logging, third-party assessments, and compliance checks so that models can be held accountable under regulations. For example, an enterprise using an LLM for content generation might need an audit trail to prove the model didn’t leak personal data or produce legally sensitive outputs, aligning with emerging AI governance frameworks \cite{M_kander_2023}.

One promising direction is the integration of cryptographic verification methods to audit model behavior without exposing sensitive details. Zero-knowledge proofs (ZKPs) allow a model owner to prove statements about the model or its outputs without revealing the underlying data. Prototype systems like \textit{ZKROWNN} have demonstrated that zk-SNARK proofs can verify a neural network’s ownership watermark efficiently, enabling IP protection for models with minimal disclosure. Similarly, the \textit{RoSeMary} framework watermarks code generated by an LLM and uses ZKPs so a third-party arbiter can confirm the code’s origin without the owner revealing the hidden signature. These approaches hint at future auditing tools where model developers can cryptographically attest to the provenance or integrity of AI-generated content. Beyond ZKPs, other privacy-preserving techniques (such as secure multiparty computation) could allow external auditors to evaluate a model on confidential data jointly with the provider - for instance, validating that a model doesn’t contain certain toxic knowledge - without either party fully seeing the other’s secrets. Such cryptographic and distributed methods could form a backbone for legal verification, giving regulators and courts verifiable evidence of a model’s compliance with IP rights, data privacy laws, or safety standards.

Another potential avenue is automated detection of unauthorized usage of LLM for malicious applications such as social media bots \cite{botLLM}. Instead of solely relying on post-hoc human review, future systems may include classifiers or filters that flag or block content flagged as machine produced. Techniques like watermarking may be further used in these contexts to filter out artificially produced content online.

\subsubsection{Sensitive Domains}

Another essential application of private computation in the domain of LLMs revolves around LLM usage on or for sensitive information. Some clear examples of this are in the fields of healthcare, biometric, or user data. The use of LLMs in healthcare has been explored for a number of possibilities such as drafting clinical reports, summarizing health records, or even going as far as making diagnostic decisions based on patient information \cite{Gencer2025-yt, mumtaz2024llmshealthcarecurrentapplications}. Google's Med-PaLM has already reached a near expert level of performance on benchmarks for medical question-answering, suggesting that AI agents may exist in the near future as assistants for doctors and patients \cite{singhal2023expertlevelmedicalquestionanswering}. However, leakage of patient data would violate strict privacy laws such as HIPAA in the US or GDPR in Europe \cite{HHS_HIPAA_Privacy_Rule_2025, GDPR_official}. Other sensitive settings such as in biometrics or finance also demand the highest standards of privacy while maintaining model accuracy \cite{pradel2021biometric, basu2021financeprivacy}. Any system leveraging LLMs to analyze biometric identifiers such as voice transcripts, fingerprints, or facial descriptions must be able to appropriately overcome legal challenges and offer ethical assurances. Laws such as Illinois' Biometric Information Privacy Act (BIPA) strictly regulate the collection and usage of biometric data \cite{BIPA_2008}.

\section*{Conclusion} 

In this work, we show how careful co-design enables privacy-preserving primitives, such as MPC, FHE, and ZKPs, to scale to the demands of LLM applications. By combining advances in hybrid protocols, leveraging quantization, automating circuit generation, and applying crypto–ML co-design, we demonstrate that secure and efficient LLM inference, ownership verification, and usage enforcement are achievable without having to compromise privacy or performance.

\section*{Acknowledgment}

This work was supported by the Defense Advanced Research Projects Agency (DARPA) Proofs program under Grant No. HR0011-23-1-0006 and by the United States National Science Foundation through the Institute for Learning-enabled Optimization at Scale (TILOS) under Award No. 2112665.

\bibliographystyle{IEEEtran}
\bibliography{references}

\end{document}